\documentclass{article}

\usepackage{fullpage}

\usepackage[affil-it]{authblk}
\usepackage{tikz}
\usepackage[utf8]{inputenc}
\usepackage[T1]{fontenc}

\usepackage{bbm}
\usepackage{amsmath}
\usepackage{amssymb}
\usepackage{mathtools}
\usepackage{optidef}
\usepackage{hyperref}
\hypersetup{
  pdftitle  = {Via a New LP Relaxation: Stronger Performance Guarantees for WSEPT, and an LP-Based Proof of Kawaguchi \& Kyan},
  pdfauthor = {Maximilian von Aspern, Andreas S. Schulz, Eugenio Vairo},
}
\usepackage{amsthm}

\newtheorem{theorem}{Theorem}
\newtheorem{lemma}{Lemma}

\newtheorem{corollary}{Corollary}

\newcommand{\E}{\mathbb{E}}

\makeatletter
\AtBeginDocument{%
  \def\doi#1{\url{https://doi.org/#1}}}
\makeatother

\usepackage{todonotes}

\begin{document}

\title{Via a New LP Relaxation:\\
Stronger Performance Guarantees for WSEPT,\\ and an LP-Based Proof of Kawaguchi \& Kyan}

\author{Maximilian {von Aspern}}
\author{Andreas S. Schulz}
\author{Eugenio Vairo}
\affil{Technical University of Munich}
\date{}

\maketitle

\begin{abstract} 
We present a new linear programming relaxation for minimizing the total weighted completion time on identical parallel machines. Its right-hand side is supermodular, so the relaxation can be solved by a greedy algorithm. We show that the WSPT rule is within a factor $\frac{1}{2}(1 + \sqrt{2})$ of the optimal LP value. This bounds the integrality gap of the relaxation and gives a short, LP-based proof of the tight bound of Kawaguchi and Kyan on the performance of WSPT. We then turn to the stochastic version of the problem, in which processing times are independent random variables whose squared coefficients of variation are at most~$\Delta$. A variance-corrected version of the new LP relaxation is valid for all non-anticipatory policies, and we show that WSEPT is within a factor $1 + (1 + \Delta)/(3 + \sqrt{7 + 2 \Delta})$ of it for $\Delta \ge 1$, for any number of machines. This is $O(\sqrt\Delta)$, whereas all previous guarantees for this problem grow linearly in $\Delta$. Up to a constant factor, our bound matches the known $\Omega(\sqrt\Delta)$ lower bound for WSEPT. For $\Delta \le 1$, we recover the best known bounds.
\end{abstract}

\section{Introduction}

One of the most fundamental and well-studied problems in the scheduling literature is the minimization of the total weighted completion time of a given set of jobs processed in parallel on several identical machines, denoted by $P\vert \, \vert \sum w_jC_j$ in the classical three-field notation for scheduling problems \cite{Graham1979}.
We are given jobs $N = \{1, \ldots, n\}$, each with processing time $p_j > 0$ and weight $w_j > 0$, and $m$ identical machines to process them.
A schedule is an assignment of jobs to machines such that each job is processed non-preemptively by exactly one machine, and each machine processes at most one job at a time.
The completion time of job~$j$ in a schedule is denoted as $C_j$, and our goal is to find a schedule which minimizes the total weighted completion time, $\sum_{j=1}^n w_j C_j$.
Because the problem is NP-hard \cite{Bruno1974}, we are interested in finding efficient algorithms that yield provably good solutions.
We call an algorithm an $\alpha$-approximation if it runs in polynomial time and produces, for any instance, a schedule whose cost is at most a factor~$\alpha$ away from the cost of an optimum. In our analyses, we actually compare the value of the algorithmic solution to the optimal value of an LP lower bound on the optimum, thereby also bounding the integrality gap of the LP. 
Since our LPs use completion-time variables, integrality is not the relevant property for defining schedules. Consequently, we henceforth use the term ``relaxation gap'' for LPs in completion-time variables instead of integrality gap, when we bound the ratio between the value of an optimal schedule and that of the LP relaxation.

\smallskip

\noindent \textbf{WSPT and the Kawaguchi \& Kyan bound:} There exist several approximation algorithms for the problem $P\vert \, \vert \sum w_jC_j$ in the literature, including a PTAS \cite{Skutella2000}. However, a simple combinatorial list scheduling algorithm, commonly referred to as WSPT (for ``Weighted Shortest Processing Time''), has arguably received most attention. WSPT assigns jobs in non-increasing order of weight-to-processing-time ratios (also known as Smith ratios \cite{Smith1956}) to the first available machine. In a landmark paper from 1986, Kawaguchi and Kyan \cite{Kawaguchi1986} showed that WSPT is a $\frac{1}{2}(1 + \sqrt{2})$-approximation algorithm, and this bound is tight.  
Some 25 years later, Schwiegelshohn \cite{Schwiegelshohn2011}, echoing community sentiment that the original proof is rather long and involved, gave a less complicated proof by reducing to sets of simple worst-case instances. Subsequently, Jäger and Skutella~\cite{Jaeger2018} refined his analysis and determined the exact performance guarantee of WSPT for every fixed number of machines. They also noted that, in contrast to most approximation results, the analysis of Kawaguchi and Kyan (and of Schwiegelshohn, and therefore their own) does not rely on a lower bound. Explicitly providing such a lower bound is the starting point of this paper.

\smallskip

\noindent \textbf{Polyhedral approaches to scheduling:} Considerable attention has been devoted to the pursuit of finding polyhedral relaxations of machine scheduling problems that can be used to establish approximation guarantees (see, e.g., \cite{Queyranne1994}). Naturally, such approaches depend more on the structure of the problem than on that of its worst-case instances. The earliest lower bound for $P\vert \, \vert \sum w_jC_j$, stated in combinatorial form, is due to Eastman, Even, and Isaacs \cite{Eastman1964}, and implies a performance guarantee of $3/2-1/(2m)$ for WSPT. It can be (re)written as 
\begin{equation}\label{oldLP}
    \min \sum_{j \in N} w_jC_j \text{ s.t. }\sum_{j \in S} p_j C_j \ge \frac{1}{2} \sum_{j \in S} p_j^2 + \frac{1}{2m} \Big (\sum_{j \in S} p_j \Big)^2 \text{ for all } S \subseteq N,
\end{equation}
see \cite{Hall1997,Queyranne1994,Schulz1996}. We are not aware of an analysis of WSPT against an LP relaxation that attains the Kawaguchi \& Kyan bound. Relaxation \eqref{oldLP} is too weak for this purpose, even if the inequalities $C_j \ge p_j$ were added for all jobs~$j$.
Chan, Muriel, and Simchi-Levi \cite{Chan1998} considered two LP relaxations---one in time-indexed variables and one in configuration variables---and established an upper bound of $\frac{1}{2}\left(1 + \sqrt{2}\right)$ on their integrality gaps. However, they used the result of Kawaguchi and Kyan to prove these bounds. More recently, Kashaev \cite{Kashaev2026} and Feng, Hu, and Li \cite{Feng2026} presented a semidefinite programming relaxation and a convex programming relaxation in assignment variables, respectively, and used these relaxations to derive the Kawaguchi \& Kyan bound. Both used the first few steps of Schwiegelshohn's \cite{Schwiegelshohn2011} `instance-reduction' and analyzed these reduced instances via their relaxations. Moreover, Kalaitzis, Svensson, and Tarnawski~\cite{Kalaitzis2017} showed that rounding the configuration LP of Sviridenko and Wiese~\cite{Sviridenko2013} is a $\frac{1}{2}\left(1 + \sqrt{2} + \epsilon \right)$-approximation for unrelated machines when all jobs have the same Smith ratio. 
% The configuration LP is at least as strong as the relaxation introduced here, so our result also bounds WSPT against it.

\smallskip

\noindent \textbf{Stochastic scheduling and WSEPT:} A stochastic version of the problem, in which processing times are drawn from independent random variables $P_1, \ldots, P_n$ whose realizations become known only upon completion of the job, has also been studied extensively.
It is customary to assume that the squared coefficients of variation of these random variables are upper-bounded by $\Delta$, i.e., $\textup{Var}[P_j]/\E[P_j]^2 \le \Delta$ for all $j \in N$.
A solution to this problem is no longer a static schedule, but an adaptive, non-anticipatory scheduling policy.
Such a policy assigns jobs non-preemptively to machines based only on information about which jobs have already been completed and which are currently being processed.
Note that these policies do not have access to the unrealized processing times of uncompleted jobs, though they are allowed to access information about their distributions.
The WSEPT policy (for ``Weighted Shortest Expected Processing Time'') is a natural extension of WSPT to the stochastic problem.
It schedules jobs in non-increasing order of $w_j/\E[P_j]$.
Note that WSEPT only requires information about the first moments of the processing time variables, rather than full distributional information.
Like its deterministic counterpart, WSEPT is optimal on a single machine but can produce suboptimal solutions on $m \ge 2$ machines \cite{Rothkopf1966}.
Möhring, Schulz, and Uetz \cite{Moehring1999} showed that WSEPT has a performance guarantee of $1 + \frac{1}{2}\left(1 + \Delta\right)$ by analyzing a stochastic extension of LP~\eqref{oldLP}.
Jäger and Skutella \cite{Jaeger2018} improved this bound  to $1 + \frac{1}{2} \frac{1 + \Delta}{1 + \min\{2, \sqrt{2+2\Delta}\}}$ by showing that the deterministic results for WSPT in \cite{Kawaguchi1986,Schwiegelshohn2011} translate into approximation guarantees for WSEPT.
This bound is still $O(\Delta)$, and no policies with performance guarantees sublinear in $\Delta$ were known prior to this paper. 
There exist lower bounds on the approximation ratio of WSEPT of $\Omega({\Delta}^{1/3})$ by Im, Moseley, and Pruhs \cite{Im2015}, and of $\Omega(\sqrt{\Delta})$ by Cheung, Fischer, Matuschke, and Megow \cite{Cheung2014}.
Furthermore, Eberle, Fischer, Matuschke, and Megow \cite{Eberle2019} gave a lower bound of $\Omega({\Delta}^{1/4})$ on the (distribution-dependent) approximation ratio of every index policy, and Skutella, Sviridenko, and Uetz \cite{Skutella2016} gave a lower bound of $\Omega(\Delta)$ for any policy that fixes all job-to-machine assignments at time $0$. 
These lower bounds even hold for the unweighted problem %(i.e., $w_j = 1$ for all $j \in N$) 
with processing times following one of two different Bernoulli distributions.
Policies with distribution-independent approximation guarantees are currently known only for the unweighted problem \cite{Antoniadis2025,Eberle2019,Gupta2023,Im2015}. It remains a major open question whether such policies exist for the weighted problem.

\smallskip

\noindent \textbf{New results:} Our contribution is threefold. 

First, we present a new LP relaxation in completion-time variables with a family of supermodular constraints that is stronger than \eqref{oldLP}, and a stochastic generalization thereof.

Second, we show that this LP relaxation yields an LP-based proof of the Kawaguchi \& Kyan bound.
Our approach differs from previous approaches in that our proof is entirely self-contained and does not rely on the techniques of Kawaguchi and Kyan \cite{Kawaguchi1986} or Schwiegelshohn \cite{Schwiegelshohn2011}, providing the first LP-based proof of this result.

Third, we improve the upper bound on the performance guarantee for WSEPT to $O(\sqrt{\Delta})$ by using the new LP relaxation, matching the lower bound up to constants.

The rest of the paper is organized as follows.
Section~\ref{sec:formulation} contains the new LP relaxation and an important auxiliary lemma.
In Section~\ref{sec:bounds}, we analyze the relaxation gap of the LP relaxation, resulting in the LP-based proof of the Kawaguchi \& Kyan bound.
In Section~\ref{sec:WSEPT}, we state and analyze the stochastic LP and give improved bounds on the performance guarantee of WSEPT.
Section~\ref{section:supermodularity} contains some further results on the structure of the new LP relaxation.

\section{A New LP Relaxation for \texorpdfstring{$P\vert \, \vert\sum w_j C_j$}{Identical Parallel Machine Scheduling}}
\label{sec:formulation}
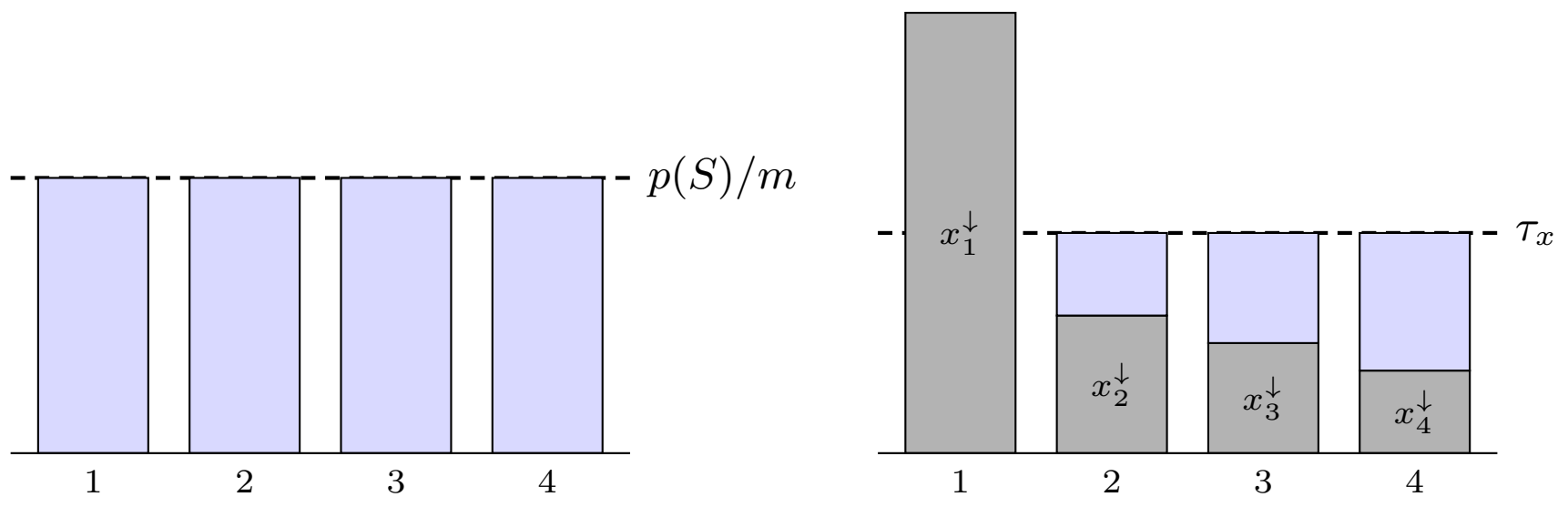
\begin{figure}[t]
\centering
\begin{tikzpicture}[x=1cm, y=0.2cm,
    job/.style={draw, fill=black!30},
    split/.style={draw, fill=blue!15},
    level/.style={densely dashed, thick},
    lab/.style={font=\scriptsize}]

  \begin{scope}
    \draw[level] (-0.2,10) -- (4.3,10) node[right, font=\small] {$p(S)/m$};
    \foreach \x in {0, 1.1, 2.2, 3.3}
      \draw[split] (\x,0) rectangle ++(0.8,10);
    \draw (-0.2,0) -- (4.3,0);
    \foreach \x [count=\i] in {0, 1.1, 2.2, 3.3}
      \node[lab, below] at (\x+0.4,0) {$\i$};
    \node[font=\small, below] at (2.05,-2.6) {(a) All jobs split};
  \end{scope}

  \begin{scope}[xshift=6.3cm]
    \draw[level] (-0.2,8) -- (4.3,8) node[right, font=\small] {$\tau_x$};
    \draw[job] (0,0)   rectangle (0.8,16);
    \draw[job] (1.1,0) rectangle (1.9,5);
    \draw[job] (2.2,0) rectangle (3.0,4);
    \draw[job] (3.3,0) rectangle (4.1,3);
    \draw[split] (1.1,5) rectangle (1.9,8);
    \draw[split] (2.2,4) rectangle (3.0,8);
    \draw[split] (3.3,3) rectangle (4.1,8);
    \node[lab] at (0.4,8)   {$x^{\downarrow}_1$};
    \node[lab] at (1.5,2.5) {$x^{\downarrow}_2$};
    \node[lab] at (2.6,2)   {$x^{\downarrow}_3$};
    \node[lab] at (3.7,1.5) {$x^{\downarrow}_4$};
    \draw (-0.2,0) -- (4.3,0);
    \foreach \x [count=\i] in {0, 1.1, 2.2, 3.3}
      \node[lab, below] at (\x+0.4,0) {$\i$};
    \node[font=\small, below] at (2.05,-2.6) {(b) Water-filling in $g(p_S)$};
  \end{scope}
\end{tikzpicture}
\caption{Two lower bounds on $\sum_{i=1}^m L_i^2(S)$ for $m=4$ machines and a job set $S$ with processing times $16,5,4,3$ and six jobs of length~$2$, where
$x=p_S$. (a)~If all jobs may be split, every load equals $p(S)/m=10$, which gives the bound $p(S)^2/m=400$ underlying~\eqref{oldLP}. (b)~In $g(p_S)$, the
$m$ longest jobs (dark) stay intact, and the remaining jobs (light) are split to raise the least-loaded machines to the water level $\tau_x=8$ (defined formally in~\eqref{eq:tau-identity-2} in Section~\ref{section:supermodularity}); machine~1 receives nothing. This gives $g(p_S)=16^2+3\cdot8^2=448$. (Unlike the usual Gantt charts, machines are drawn as columns, with time running upward, to emphasize the water-filling analogy.)}
\label{fig:waterfilling}
\end{figure}

In this section, we present a new linear programming relaxation of $P\lvert \, \rvert \sum w_j C_j$. 
For any instance, it is at least as strong as relaxation \eqref{oldLP}. 
In the following, we assume that $n > m \ge 2$; otherwise, the problem is optimally solved by the WSPT algorithm.
Consider the functions $g: \mathbb{R}^n_{\ge 0} \to \mathbb{R}_{\ge 0}$ and $f: 2^{N} \to \mathbb{R}_{\ge 0}$ defined via
\begin{align*}
    g(x) &= \min \Big \{\sum_{i=1}^m L_i^2 : \sum_{i=1}^m L_i = \sum_{j=1}^n x_j, L_i \ge x^\downarrow_{i} \ \text{ for all } i \in \{1,\ldots,m\} \Big \},\\
    f(S) &= \frac{1}{2} p^2(S) + \frac{1}{2} g(p_S).
\end{align*}
Here, $x^\downarrow_{i}$ denotes the $i$-th largest entry of the vector $x \in \mathbb{R}^n_{\ge 0}$, and $p_S \in \mathbb{R}^n_{\ge 0}$ is a vector representation of the processing times in $S \subseteq N$ with its $j$-th entry equal to $p_j$ if $j \in S$, and $0$ otherwise, i.e., $p_S = (p_j \mathbbm{1}_{j\in S})$. 
This is not to be confused with the notation $p(S) = \sum_{j \in S} p_j$ and $p^2(S) = \sum_{j \in S} p_j^2$, which we use throughout the paper.
The LP relaxation can now be stated as follows:
\begin{mini!}
{}{\sum_{j=1}^n w_j C_j}
{\label{LP}}{}
\addConstraint{\sum_{j \in S} p_j C_j}{\ge f(S)}{\quad \text{ for all } S \subseteq N. \label{water-filling inequalities}}
\end{mini!}
The formulation of the constraint family~\eqref{water-filling inequalities}, and specifically of the function~$g$, uses water-filling ideas that have previously been explored when analyzing $P\vert \, \vert\sum w_jC_j$ and WSPT, see \cite{Jaeger2018,Kashaev2026,Schwiegelshohn2011}, albeit not to formulate an LP relaxation.
It is well known and easy to verify that for any non-idling schedule of jobs in $S$, 
\begin{equation}
\label{eq:load-identity}
\sum_{j\in S} p_jC_j = \frac{1}{2}p^2(S) + \frac{1}{2} \sum_{i=1}^m L_i^2(S),
\end{equation}
where $L_i(S)$ denotes the load of machine $i$, i.e., the total processing time of jobs in $S$ assigned to it.
Because the first term is constant, we will ignore it for now. 
If jobs could be split arbitrarily among the machines, the second term would be
smallest when all loads are equal, that is,
$$
  \sum_{i=1}^m L_i^2(S)\ \ge\ \min\Big \{\sum_{i=1}^m L_i^2 :
  \sum_{i=1}^m L_i = p(S)\Big \}\ =\ \frac{1}{m}\,p(S)^2 .
$$
This is the bound of Eastman, Even, and Isaacs~\cite{Eastman1964} that underlies
LP~\eqref{oldLP}. It ignores that jobs cannot be split and is therefore too
optimistic when some jobs are long. For instance, for $S=\{j\}$,
\eqref{oldLP} only yields $C_j\ge\frac{1}2(1+\frac{1}m)\,p_j$, whereas every
schedule has $C_j\ge p_j$. The new relaxation keeps the $m$ longest jobs intact:
the machine that processes a job has load at least the job's length, and in an
assignment that minimizes $\sum_i L_i^2(S)$, the $i$-th largest load is at
least the $i$-th largest processing time in~$S$.
(Theorem~\ref{theorem:LP-vailidity} makes this precise.) Adding these $m$ lower bounds
to the minimization above yields exactly $g(p_S)$. Only the remaining jobs are
split, and not evenly. In an optimal solution, they raise the least-loaded
machines to a common level, and machines whose long job already exceeds this
level receive nothing. This is the water-filling structure of~$g$ (see Figure~\ref{fig:waterfilling}).

Let us now formally prove the validity of the new relaxation.

\begin{theorem}
\label{theorem:LP-vailidity}
    For every feasible schedule and every $S \subseteq N$, 
    $$\sum_{j \in S} p_j C_j \ge f(S).$$
\end{theorem}

\begin{proof}
Delete the jobs in $N\setminus S$ and shift the remaining jobs on each machine to the left, so that they run without idle time from time~$0$; denote the new completion times by $C^S_j$. This does not increase any completion time, so by \eqref{eq:load-identity}
\begin{equation*}
    \sum_{j \in S} p_j C_j \ge \sum_{j \in S} p_j C^S_j = \frac{1}{2} p^2(S) + \frac{1}{2}\sum_{i=1}^m L_i^2(S).
\end{equation*}
Here, $L_i(S)$ denotes the total processing time of jobs in $S$ assigned to machine~$i$. \\

\noindent It therefore suffices to show $\sum_{i=1}^m L_i^2(S) \ge g(x)$ for $x=p_S$ and every assignment of the jobs of $S$ to the machines. Among the finitely many assignments, choose one that minimizes $\sum_{i=1}^m L_i^2(S)$. Without loss of generality, we may assume $L_1(S)\ge\ldots\ge L_m(S)$. If $L_i(S) \ge x_i^\downarrow$ for all $i$, then $\sum_{i=1}^m L_i^2(S) \ge g(x)$ by the definition of $g$.\\

\noindent Suppose instead that $L_k(S) < x_k^\downarrow$ for some $k\le m$. Because the schedule must not preempt the jobs of length $x_1^\downarrow, \ldots, x_k^\downarrow$, there also exists a machine $h < k$ which schedules two jobs of length at least $x_k^\downarrow$. Choose one of these jobs arbitrarily and denote its length by $y \ge x_k^\downarrow$. Exchanging this job with all jobs scheduled on machine $k$ decreases $\sum_{i=1}^m L_i^2(S)$ by
$$
L_h^2(S) + L_k^2(S) - \big(L_h(S) + L_k(S) - y \big)^2 - y^2 = 2\big(L_h(S)-y\big)\big(y-L_k(S)\big). 
$$
Note that $y-L_k(S)\ge x^\downarrow_k-L_k(S)>0$, and $L_h(S)-y>0$, because machine $h$ also holds the second job, whose length is at least $x^\downarrow_k>0$. Hence, the decrease is strictly positive, which contradicts the choice of the assignment.
\end{proof}

As alluded to before, it is easy to see that the constraints \eqref{water-filling inequalities} are at least as strong as the constraints in LP \eqref{oldLP}. Furthermore, the new constraints also include $C_j \ge p_j$, which can be observed by noting that $f(\{j\}) = g(p_{\{j\}}) = p_j^2$.

The following auxiliary result is crucial to our analysis in the following sections. Throughout the paper, we write $(y)_+=\max\{y,0\}$ and $(y)_+^2=((y)_+)^2$.

\begin{lemma}\label{lemma:lb_g}
Consider a schedule of a job set $S$. Let $\ell_i$ be the processing time of any one job on machine~$i$ (and $\ell_i = 0$ for an empty machine). Let $L_i$ be the total processing time of all jobs assigned to machine~$i$. Then, for every $\nu\ge0$,
$$
  \sum_{i=1}^m\Big(2\nu L_i-\nu^2+ (\ell_i-\nu)_+^2\Big) \le g(p_S).
$$
\end{lemma}

\begin{proof}
Let $L^*$ attain the minimum in the definition of $g$ for $x = p_S$.  Because $L^*_i \ge x^{\downarrow}_i$, it holds that $(L^*_i - \nu)^2 \ge (x^{\downarrow}_i - \nu)^2_+$, and with $\sum_{i=1}^m L_i = \sum_{i=1}^m L^*_i = p(S)$, we get
\begin{equation}
    \label{eq:lb_g}
    g(p_S) = \sum_{i=1}^m (L^*_i - \nu)^2 + 2 \nu \sum_{i=1}^m L_i - m \nu^2 \ge \sum_{i=1}^m (x^{\downarrow}_i - \nu)^2_+ + 2 \nu \sum_{i=1}^m L_i -  m \nu^2. 
\end{equation}
The nonzero $\ell_i$ are processing times of distinct jobs of $S$; and therefore $\sum_{i=1}^m (x^{\downarrow}_i - \nu)^2_+ \ge \sum_{i=1}^m (\ell_i - \nu)^2_+$.
\end{proof}

\section{Bounding the Relaxation Gap and an LP-Based Proof of the Kawaguchi \texorpdfstring{\&}{&} Kyan Bound}
\label{sec:bounds}

To bound the relaxation gap of \eqref{LP}, we want to find an algorithm $\mathcal{A}$ and an $\alpha \ge 1$ such that, for all instances, $\sum_{j=1}^n w_j C_j^\mathcal{A} \le \alpha \sum_{j=1}^n w_j C_j^\textup{LP}$, where $C^\textup{LP}$ denotes an optimal solution to the LP relaxation \eqref{LP}.
This would imply that $\mathcal{A}$ is an $\alpha$-approximation and that the relaxation gap of LP \eqref{LP} is bounded from above by $\alpha$.

For this purpose, we will use WSPT. We assume w.l.o.g. that the jobs are sorted by their Smith ratios, i.e., $w_1/p_1 \ge w_2/p_2 \ge \ldots \ge w_n/p_n$.
Note that it is well known that WSPT has a tight approximation guarantee of $\frac{1}{2}\left(1 + \sqrt{2}\right)$ \cite{Kawaguchi1986}. 
Therefore, 
the best bound on the relaxation gap obtainable with the choice $\mathcal{A} = \textup{WSPT}$ is exactly this factor, and indeed, we will show that this bound can be achieved.
This result provides an alternative proof of the approximation guarantee of WSPT, which is arguably more compact than previous proofs and is the first LP-based proof of this result.

\begin{theorem}\label{theorem:KK}
For any instance of $P\vert \, \vert\sum w_j C_j$, the WSPT schedule satisfies
$$
\sum_{j=1}^n w_j C_j^{\rm WSPT} \quad \le \quad \frac{1}{2} \Big(1+\sqrt{2} \Big) \, \cdot \,\sum_{j=1}^n w_j C_j^{\rm LP}.
$$
\end{theorem}

\begin{proof}
Let $\rho = \frac{1+\sqrt{2}}{2}$. 
With $\frac{w_{n+1}}{p_{n+1}} = 0$, the cost of the WSPT schedule can be rewritten as
$$
\sum_{j=1}^n w_j C_j^{\rm WSPT} = \sum_{k=1}^n \left(\frac{w_k}{p_k} - \frac{w_{k+1}}{p_{k+1}}\right) \sum_{j=1}^k p_j C_j^{\rm WSPT}.
$$
Because $w_k/p_k \ge  w_{k+1}/p_{k+1}$ for all $k \in N$, it suffices to show, for every $k \in N$, that the set $S = \{1,\ldots,k\}$ satisfies $\sum\limits_{j \in S}p_jC_j^{\rm WSPT} \le \rho \cdot f(S)$. The latter inequality is equivalent to 
\begin{equation}\label{eq:load^2}
\sum_{i=1}^m L_i^2 \le \rho \cdot g(p_S) + (\rho -1 ) \cdot p^2(S) ,
\end{equation}
where $L_i$ is the corresponding load on machine $i$.
So let $S$ be such a set, and let $\eta = \min_i L_i$.
Note that no job can be started by WSPT after $\eta$ because a machine becomes idle from time $\eta$ on, so WSPT would have scheduled that job at time $\eta$.\\

\noindent If $\eta = 0$, then every job starts at time $0$.  Hence, each machine processes at most one job and, therefore, $\sum_{i=1}^m L_i^2 = p^2(S) = g(p_S)$, and the inequality holds.\\

\noindent Otherwise, we scale the processing times so that $\eta = 1$.  For each machine $i$, let $\ell_i$ be the length of its last job; its start time $L_i-\ell_i$ is at most $\eta$. Each machine then satisfies $1 \le L_i \le 1 + \ell_i$.
With Lemma~\ref{lemma:lb_g} and with $\sum_{i=1}^m \ell_i^2 \le p^2(S)$, it suffices to prove, for each machine~$i$, that
\begin{equation*}
L_i^2 \le \rho \Big(2 \nu L_i - \nu^2 + (\ell_i - \nu)^2_+\Big) + (\rho-1) \ell_i^2 .
\end{equation*}
With $\rho = \frac{1+\sqrt{2}}{2}$ and $\nu = \sqrt{2}$, this is equivalent to showing that
$$
(L_i-1) (1 + \sqrt{2}-L_i) + \frac{\sqrt{2}-1}{2} \ell_i^2 + \frac{\sqrt{2}+1}{2} (\ell_i  - \sqrt{2})^2_+ \ge 0.
$$
For $L_i \le 1 + \sqrt{2}$, all three terms are nonnegative, so the claim is trivial.\\

\noindent So let's assume $L_i > 1 + \sqrt{2}$. With our earlier bound, we have $\ell_i \ge L_i -1 > \sqrt{2}$. The function $x \mapsto x(\sqrt{2}-x)$ is decreasing for $x > \sqrt{2}$. Together, this implies that $(L_i-1) (1 + \sqrt{2}-L_i) \ge \ell_i (\sqrt{2} - \ell_i)$ and we have

\begin{align*}
    (L_i-1) (1 + \sqrt{2}-L_i) + \frac{\sqrt{2}-1}{2} \ell_i^2 + \frac{\sqrt{2}+1}{2} (\ell_i  - \sqrt{2})^2_+ & \ge \\
    \ell_i (\sqrt{2} - \ell_i) + \frac{\sqrt{2}-1}{2} \ell_i^2 + \frac{\sqrt{2}+1}{2} (\ell_i  - \sqrt{2})^2 & = \\
    (\sqrt{2}-1) (\ell_i - \sqrt{2} - 1)^2 & \ge 0.
\end{align*}
\end{proof}

Even though the Kawaguchi \& Kyan bound is tight as an approximation guarantee for WSPT, the same need not hold for the relaxation gap of LP~\eqref{LP}. Indeed, the exact value of the relaxation gap remains open. 
We provide a lower bound of $9/8$.
Consider an instance with $m=2$ machines and three jobs with processing times $1,1,3$ and unit weights.
By the optimality of SPT for $P\vert \, \vert\sum C_j$~\cite{Conway1967}, the optimal solution is $C^*_1 = C^*_2 = 1$ and $C^*_3 = 4$.
Meanwhile, an optimal LP solution is $C^\textup{LP}_1 = C^\textup{LP}_2 = 1$ and $C^\textup{LP}_3 = 10/3$.
Together, this yields
\begin{equation*}
    \frac{C^*_1 + C^*_2 + C^*_3}{C^\textup{LP}_1 + C^\textup{LP}_2 + C^\textup{LP}_3} = \frac{6}{16/3} = \frac{9}{8}.
\end{equation*}

\section{An \texorpdfstring{$O(\sqrt\Delta)$}{O(\sqrt\Delta)} Approximation Guarantee for WSEPT}
\label{sec:WSEPT}

The main result in this section is an improved performance guarantee of WSEPT. 
Our analysis recovers the bound of Jäger and Skutella \cite{Jaeger2018} for $\Delta \le 1$ but strictly improves upon their $1 + \frac{1+\Delta}{6}$ bound for $\Delta > 1$.

\begin{theorem}
\label{theorem:WSEPT}
The performance guarantee of WSEPT is at most
$$
1 +
\begin{cases}
\frac{1+\Delta}{2 + \sqrt{8(1+\Delta)}} & \text{if } \Delta \le 1 , \\[0.3cm]
\frac{1+\Delta}{3 + \sqrt{7+2\Delta}} & \text{if } \Delta \ge 1 .
\end{cases}
$$
In particular, it grows like $\sqrt{\Delta/2}$ as $\Delta \rightarrow \infty$.
\end{theorem}
To prove this result, we combine a refined version of the analysis from Section~\ref{sec:bounds} with a stochastic extension of LP~\eqref{LP}.

\subsection{A Refined Analysis of the LP Relaxation}
The key step in Theorem~\ref{theorem:KK} was to prove inequality \eqref{eq:load^2}.  In the proof, we chose $\rho$ and $\nu$ to get the Kawaguchi \& Kyan bound, which is tight.  However, we could have continued the proof with any $\sqrt{2} \le \nu \le 3/2$ and for any list schedule, and would have obtained the following lemma.
\begin{lemma}\label{remark: refined analysis}
Let $\sqrt{2} \le \nu \le 3/2$. Then every list schedule of every set $S$ of jobs satisfies
\begin{equation}\label{eq:general_bound_on_load^2}
\sum_{i=1}^m L_i^2 \le \frac{1}{\nu(2-\nu)} \cdot g(p_S) + \frac{3 - 2 \nu}{\nu (2 - \nu)} \cdot p^2(S) .
\end{equation}
\end{lemma}
\begin{proof}
The proof of this statement is analogous to that of Theorem~\ref{theorem:KK}. As before, let $\eta = \min_i L_i$. 
If $\eta = 0$, the claim reduces to $2 \ge \nu$, which is true.\\

\noindent Otherwise, we scale the processing times so that $\eta = 1$.  Each machine then satisfies $1 \le L_i \le 1 + \ell_i$.
With the bound from Lemma~\ref{lemma:lb_g} and with $\sum_{i=1}^m \ell_i^2 \le p^2(S)$, it suffices to prove, for each machine~$i$, that
\begin{equation*}
L_i^2 \le \frac{1}{\nu(2-\nu)} \Big(2 \nu L_i - \nu^2 + (\ell_i - \nu)^2_+\Big) + \frac{3 - 2 \nu}{\nu (2 - \nu)} \ell_i^2 .
\end{equation*}
This is equivalent to showing that
$$
(L_i-1) (\frac{\nu}{2-\nu}-L_i) + \frac{3 - 2 \nu}{\nu (2 - \nu)} \ell_i^2 + \frac{1}{\nu(2-\nu)} (\ell_i  - \nu)^2_+ \ge 0.
$$
For $L_i \le \frac{\nu}{2-\nu}$, all three terms are nonnegative, so the claim is trivial.\\

\noindent So let's assume $L_i > \frac{\nu}{2-\nu}$. Then we have $\ell_i \ge L_i -1 > \frac{2(\nu-1)}{2-\nu}\ge \nu$ because $\nu \ge \sqrt{2}$. The function $x \mapsto x(\frac{2(\nu-1)}{2-\nu}-x)$ is decreasing for $x > \frac{2(\nu-1)}{2-\nu}$. Together, this implies that $(L_i-1) (\frac{\nu}{2-\nu}-L_i) \ge \ell_i (\frac{2(\nu-1)}{2-\nu} - \ell_i)$ and we have
\begin{align*}
    (L_i-1) (\frac{\nu}{2-\nu}-L_i) + \frac{3 - 2 \nu}{\nu (2 - \nu)} \ell_i^2 + \frac{1}{\nu(2-\nu)} (\ell_i  - \nu)^2_+ & \ge \\
    \ell_i (\frac{2(\nu-1)}{2-\nu} - \ell_i) + \frac{3 - 2 \nu}{\nu (2 - \nu)} \ell_i^2 + \frac{1}{\nu(2-\nu)} (\ell_i  - \nu)^2 & = \\
    \frac{2-\nu}{\nu} (\ell_i - \frac{\nu}{2-\nu})^2 & \ge 0.
\end{align*}
\end{proof}
We use this more general bound on the squared machine loads to prove the bounds in Theorem~\ref{theorem:WSEPT} for $\Delta \le 1$. To obtain our results for $\Delta > 1$ and thus to prove better performance guarantees for WSEPT, we complement it with the following lemma.

\begin{lemma}
Let $1 < \nu \le \frac{3}{2}$. Then every list schedule of every set $S$ of jobs satisfies
\begin{equation}\label{eq:general_bound_on_load^2_no2}
\sum_{i=1}^m L_i^2 \le \frac{1}{\nu(\nu-1)} \cdot g(p_S) - \frac{3 - 2 \nu}{\nu -1} \cdot p^2(S) .
\end{equation}
\end{lemma}

\begin{proof}
As in the proof of Theorem~\ref{theorem:KK}, let $\eta = \min_i L_i$. If $\eta = 0$, the claim reduces to $(\nu-1)/\nu \ge 0$, which is true.\\
    
\noindent So let's consider the case $\eta > 0$ and again scale the processing times so that $\eta =1$. We use $s_i = L_i - \ell_i$ to denote the start time of the last job on machine $i$. Hence, $0 \le s_i \le 1 \le L_i$. \\

 \noindent With $p^2(S) \le \sum_{i=1}^m(\ell^2_i + s^2_i) = \sum_{i=1}^m(L_i^2 - 2 s_i \ell_i)$ and with the lower bound from Lemma~\ref{lemma:lb_g}, it suffices to prove, for each machine~$i$, 
 $$
 \nu (L_i-1) (\nu - (2-\nu)L_i)+2\nu(3-2\nu)s_i\ell_i+(\ell_i-\nu)^2_+ \ge 0 .
 $$
 Using $L_i = s_i + \ell_i$, this is equivalent to
 $$
 \big ( (\nu -1)\ell_i-\nu s_i \big )^2 + 2 \nu s_i(1-s_i) - (\nu - \ell_i)^2_+ \ge 0 .
 $$
If $\ell_i \ge \nu$, the last term is zero, so the inequality holds.\\

\noindent If $\ell_i \le \nu$, then, for fixed $s_i$, the left-hand side is concave in $\ell_i$. Therefore, it suffices to check the endpoints of $\ell_i \in [1-s_i,\nu]$. At $\ell_i = \nu$, we know it is greater than or equal to zero, by the first case. Plugging in $\ell_i = 1 - s_i$ simplifies the left-hand side to $2 \nu (3-2\nu)s_i(1-s_i)$, where all factors are nonnegative.
\end{proof}

\subsection{A Stochastic Generalization of the LP Relaxation}
\label{sec:stochastic LP}

We give a version of LP \eqref{LP} that applies to stochastic scheduling.
In the following, we use $\mu_j$ to denote the expected processing times, and $\sigma^2_j$ to denote their variances. We assume $\sigma_j^2 \le \Delta \, \mu_j^2$ for all jobs $j$.
Note that WSEPT requires only the expected processing times and no additional distributional information.
In particular, we need not know the variances or $\Delta$ to run WSEPT; we only use them in the analysis.
We use $\mu^2(S) = \sum_{j \in S} \mu_j^2$ and $\sigma^2(S) = \sum_{j \in S} \sigma_j^2$. Let $P_S$ be the random vector with entries $P_j$ for $j\in S$, and $0$ otherwise. Then $g(P_S)$ is the random variable obtained by applying $g$ to the realized processing times, and $\E[g(P_S)]$ denotes its expectation.\footnote{Note that this expectation is well defined and finite. In general, it depends on the marginal distributions of all processing times, but we need not compute it to run WSEPT, nor for our analysis.}
Our approach is similar in spirit to that of Möhring, Schulz, and Uetz \cite{Moehring1999}, who generalized LP \eqref{oldLP} to the stochastic problem to obtain the first $O(\Delta)$ bound on WSEPT.
\begin{mini!}
{}{\sum_{j=1}^n w_j \E[C_j]}
{\label{stochastic LP}}{}
\addConstraint{\sum_{j \in S} \mu_j \E[C_j]}{\ge \frac{1}{2} \Big( \mu^2(S) - \sigma^2(S) + \E[g(P_S)] \Big)}{\quad \text{ for all } S \subseteq N. \label{stochastic water-filling inequalities}}
\end{mini!}

In the following, we refer to the right-hand side of constraints \eqref{stochastic water-filling inequalities} as $F(S)$.
We first prove that this is a valid relaxation.

\begin{theorem}
For every non-anticipatory policy and every $S \subseteq N$, 
$$\sum_{j \in S} \mu_j \E[C_j] \ge F(S).$$
\end{theorem}

\begin{proof}
    Consider an arbitrary non-anticipatory policy.
    For any realization of processing times and any $S \subseteq N$, it induces a feasible schedule. 
    By Theorem~\ref{theorem:LP-vailidity}, this schedule satisfies $\sum_{j \in S} p_j S_j\ge - \frac{1}{2} p^2(S) + \frac{1}{2} g(p_S) $, where $S_j \coloneqq C_j - p_j$ denotes the start time of job $j$ under the given policy and processing time realization.\\

    \noindent The start time of a job does not depend on its processing time because the policy is non-anticipatory and job processing times are independent of each other. Therefore, taking expectations on both sides yields
    $$
    \sum_{j \in S} \mu_j \E[S_j]\ge - \frac{1}{2} \E[\sum_{j \in S} P_j^2] + \frac{1}{2} \E[g(P_S)]  = -\frac{1}{2} \big( \sigma^2(S) + \mu^2(S) \big ) + \frac{1}{2} \E[g(P_S)],
    $$
    and adding $\mu^2(S)$ to both sides completes the proof.
    We have used here that $\E[\sum_{j \in S} P^2_j] = \mu^2(S) + \sigma^2(S)$.
\end{proof}

As in the deterministic case, inequalities~\eqref{stochastic water-filling inequalities} dominate the natural inequalities $\mathbb{E}[C_j]\ge\mu_j$, which they contain for $S=\{j\}$, as well as the inequalities
$$
  \sum_{j\in S}\mu_j\mathbb{E}[C_j]\ \ge\ \frac{1}2\bigl(\mu^2(S)-\sigma^2(S)\bigr)
  +\frac{1}{2m}\bigl(\mu(S)^2+\sigma^2(S)\bigr)\qquad\text{for all }S\subseteq N
$$
of Möhring, Schulz, and Uetz \cite{Moehring1999}. Indeed,
$g(x)\ge\frac{1}m\bigl(\sum_jx_j\bigr)^2$ for all $x$ by the Cauchy--Schwarz
inequality, as in the comparison with~\eqref{oldLP}, and
$\mathbb{E}\bigl[\bigl(\sum_{j\in S}P_j\bigr)^2\bigr]=\mu(S)^2+\sigma^2(S)$ by
independence.

\subsection{Proof of Theorem\texorpdfstring{~\ref{theorem:WSEPT}}{ 3}}

\begin{lemma}\label{lemma:WSEPT}

Let $a \ge 1$ and $b \ge 0$. 
Suppose that for every subset $S$ of jobs and every realization of processing times, the list schedule of $S$ in WSEPT order satisfies 
$$
\sum_{i=1}^m L_i^2 \le a \cdot g(p_S) + (1 + 2b - a) \cdot p^2(S).
$$
Then WSEPT has a performance guarantee of $\max\{a, 1 + (1+\Delta)b\}$.
\end{lemma}

\begin{proof}
The WSEPT order does not depend on the realization of processing times, but only on their means. 
So we can assume that $w_1/\mu_1 \ge \ldots \ge w_n/\mu_n$, i.e., WSEPT schedules jobs in order $1,\ldots,n$.\\

\noindent As in the deterministic case, it suffices to show, for every $S = \{1, \ldots, k\}$, that $\sum_{j \in S} \mu_j \E[C^{\rm WSEPT}_j] \le \rho \cdot F(S)$, for $\rho = \max\{a, 1 + (1+\Delta)b\}$.
We have 
\begin{equation}
\label{eq:stochastic-identity}
    F(S) - \mu^2(S) = \frac{1}{2} \E\Big[g(P_S) - \sum_{j \in S} P^2_j\Big] \ge 0 . 
\end{equation}
Here, the inequality holds because $g(x) \ge \sum_{j=1}^n x_j^2$ for all $x \in \mathbb{R}^n_{\ge 0}$. For every realization, the assumption implies
$$
\sum_{j \in S} p_jS_j = \frac{1}{2} \Big ( \sum_{i=1}^m L_i^2 - p^2(S) \Big ) \le \frac{a}{2} \Big ( g(p_S) - p^2(S) \Big ) + b \, p^2(S) .
$$
Taking expectations and using \eqref{eq:stochastic-identity}, the right-hand side becomes $a(F(S) - \mu^2(S)) + b(\mu^2(S) + \sigma^2(S))$.  Hence, we get
$$
\sum_{j \in S} \mu_j \E[C^{\rm WSEPT}_j] \le a \Big (F(S) - \mu^2(S) \Big ) + \Big (1+(1+\Delta)b \Big ) \mu^2(S) \le \rho \cdot F(S) .
$$
\end{proof}
To prove Theorem~\ref{theorem:WSEPT}, we apply Lemma~\ref{lemma:WSEPT} to inequalities \eqref{eq:general_bound_on_load^2} and \eqref{eq:general_bound_on_load^2_no2} for $\Delta\le1$ and $\Delta\ge1$, respectively.
In both cases, we choose $\nu$ so that $a = 1 + (1 + \Delta) b$.

\begin{proof}[Proof of Theorem~\ref{theorem:WSEPT}]
For $\Delta \le 1$, we note that we can rewrite the right-hand side of \eqref{eq:general_bound_on_load^2} as
$$
a \cdot g(p_S) + (1 + 2b - a) \cdot p^2(S)
$$
with $a = \frac{1}{\nu(2-\nu)}$ and $b=\frac{2-\nu}{2\nu}$ because $\nu(2-\nu) + (2-\nu)^2-1 = 3 - 2\nu$.\\

\noindent For $\Delta = 1$, the choice $\nu = \frac{3}{2}$ satisfies $a = 1 + (1 + \Delta) b$, yielding a performance guarantee of $a=\frac{1}{\nu(2-\nu)} = \frac{4}{3}$ via Lemma~\ref{lemma:WSEPT}.\\

\noindent For $\Delta < 1$, the choice $\nu = \frac{\sqrt{2(1+\Delta)} - 2\Delta}{1-\Delta} \in [\sqrt{2},\frac{3}{2}]$ satisfies $a = 1 + (1 + \Delta) b$, yielding a performance guarantee of 
$$a=\frac{1}{\nu(2-\nu)} = 1 + \frac{1+\Delta}{2+\sqrt{8(1+\Delta)}}.$$
Similarly, for $\Delta \ge 1$, we bring \eqref{eq:general_bound_on_load^2_no2} into the form required by Lemma~\ref{lemma:WSEPT} with $a = \frac{1}{\nu(\nu-1)}$ and $b = \frac{\nu-1}{2 \nu}$ because $\nu(\nu-1) + (\nu-1)^2 - 1 = \nu(2\nu-3)$.\\

\noindent The choice $\nu = \frac{2+\Delta+\sqrt{7+2\Delta}}{3+\Delta}\in (1,\frac{3}{2}]$ satisfies $a = 1 + (1 + \Delta) b$ and we get the performance guarantee
$$
a=\frac{1}{\nu(\nu-1)} = 1 + \frac{1+\Delta}{3+\sqrt{7+2\Delta}}.
$$
Note that the choices of $\nu$ clearly satisfy $a\ge1$ and $b\ge0$, which follows from $a = 1 + (1 + \Delta) b = 1 + \frac{1+\Delta}{2+\sqrt{8(1+\Delta)}}$ and $a = 1 + (1 + \Delta) b = 1 + \frac{1+\Delta}{3+\sqrt{7+2\Delta}}$, respectively.
\end{proof}

\section{Further Results}
\label{section:supermodularity}

The constraint family \eqref{water-filling inequalities} contains exponentially many constraints, which raises the question of whether one can solve relaxation \eqref{LP} in polynomial time.  Note that this is not needed for any of the results presented so far. 
Still, we show that the answer is positive by proving that $f$ is a supermodular function. In fact, this allows us to solve the LP in strongly polynomial time. 
Furthermore, the supermodularity of $f$ allows us to make additional structural observations.
Our main tool is another characterization of the water-filling function $g$, which also makes the water-filling interpretation of the new inequalities explicit.

\smallskip

We say that a set function $\varphi: 2^{N} \to \mathbb{R}$ is supermodular if $\varphi(S) + \varphi(T) \le \varphi(S \cup T) + \varphi(S\cap T)$ for all $S,T \subseteq N$. It is well known that this characterization is equivalent to the following \cite{Nemhauser1978,Topkis1978}, which we use in the remainder of this section:
\begin{equation*}
    \varphi(S \cup \{j\}) - \varphi(S) \le \varphi(T \cup \{j\}) - \varphi(T) \quad \text{for all } S \subseteq T \subseteq N \setminus \{j\}.
\end{equation*}
If $\varphi$ is supermodular, $\varphi(\emptyset)=0$, and $v \ge 0$, a supermodular LP is given by 
$$
\text{minimize } \sum_{j \in N} v_j x_j \qquad \text{subject to } \sum_{j \in S} x_j \ge \varphi(S) \quad \text{for all } S \subseteq N.
$$
The corresponding polyhedron is also known as the contrapolymatroid of $\varphi$.
If $v_1 \ge v_2 \ge \ldots \ge v_n$, Edmonds \cite{Edmonds1970} showed that $x_j^* = \varphi([j]) - \varphi([j-1])$, $j \in N$, is an optimal solution, where 
$[a] = \{1,\ldots,a\}$ for $a \in \mathbb{N}_{>0}$ and $[0] = \emptyset$. This greedy algorithm runs in strongly polynomial time if $\varphi$ can be evaluated in strongly polynomial time.

\smallskip

For $x\in\mathbb{R}^n_{\ge0}$, the ``water level'' $\tau_x$ is the unique $\tau\ge x^\downarrow_m$ with
\begin{equation}\label{eq:tau-identity-2}
  \sum_{i=1}^m\max\{x^\downarrow_i,\tau\}=\sum_{j=1}^nx_j .
\end{equation}
It exists and is unique because the left-hand side is continuous in $\tau$, equals $\sum_{j=1}^mx^\downarrow_j\le\sum_{j=1}^n x_j$ at $\tau=x^\downarrow_m$, and
is strictly increasing for $\tau\ge x^\downarrow_m$. Since $\tau_x\ge x^\downarrow_m$, all entries of $x$ that exceed $\tau_x$ are among $x^\downarrow_1,\dots,x^\downarrow_{m-1}$.

\begin{lemma}\label{lem:waterlevel}
For every $x\in\mathbb{R}^n_{\ge0}$, the vector
$L^*_i=\max\{x^\downarrow_i,\tau_x\}$, $i=1,\dots,m$, attains the minimum in the
definition of $g$. Hence
\begin{equation}\label{eq:tau-identity-1}
  g(x)=\sum_{i=1}^m\max\{x^\downarrow_i,\tau_x\}^2 .
\end{equation}
\end{lemma}

\begin{proof}
By~\eqref{eq:tau-identity-2}, $L^*$ is feasible. For every feasible $L$,
$$
  \sum_{i=1}^mL_i^2-\sum_{i=1}^m(L^*_i)^2 \ge 2\sum_{i=1}^mL^*_i(L_i-L^*_i) \ge 2\tau_x\sum_{i=1}^m(L_i-L^*_i) = 0,
$$
where the first inequality follows from $a^2-b^2=(a-b)^2+2b(a-b)$ for all $a,b\in\mathbb{R}$. The second inequality holds because either $L^*_i=\tau_x$, or $L^*_i=x^\downarrow_i>\tau_x$ and then $L_i\ge x^\downarrow_i=L^*_i$ by feasibility. The final equality holds because
$\sum_{i=1}^mL_i=\sum_{i=1}^mL^*_i=\sum_{j=1}^nx_j$. So $L^*$ is optimal.
\end{proof}

Note that if $x^\downarrow_{k-1}>\tau_x\ge x^\downarrow_k$ (with $x^\downarrow_0 =\infty$), then~\eqref{eq:tau-identity-2} is equal to $\sum_{i=1}^{k-1}x^\downarrow_i+(m-k+1)\tau_x=\sum_{j=1}^nx_j$, i.e., $\tau_x=\frac{1}{m-k+1}\sum_{j=k}^nx^\downarrow_j$. Any $k\in\{1,\dots,m\}$ for which this value lies in $[x^\downarrow_k,x^\downarrow_{k-1})$ yields a solution of~\eqref{eq:tau-identity-2} that is at least $x^\downarrow_m$, so $\tau_x$ can be computed by trying $k=1,\dots,m$. Hence $g$, and thus $f$, can be evaluated in strongly polynomial time.

\begin{lemma}\label{lemma:water-filling}
The function $g$ and the water level $\tau_x$ have the following properties for all $x\in\mathbb{R}_{\ge0}^n$ and all $j\in N$:
\begin{enumerate}
  \item $\tau_x$ is continuous and non-decreasing in $x_j$, and
  \item $\frac{\partial}{\partial x_j}g(x)=2\max\{x_j,\tau_x\}$ for $x_j > 0$.
\end{enumerate}
\end{lemma}

\begin{proof}
For arbitrary $x\in\mathbb{R}_{\ge0}^n$, $j\in N$, $t \ge 0$, let $x(t)$ be the vector with $x_j(t) = t$ and $x_l(t) = x_l$ for $l\neq j$. Furthermore, for all $r\in\{0,\ldots,m-1\}$, let
$$
T_r(t):=\frac{1}{m-r}\sum_{i=r+1}^n x_i^\downarrow(t).
$$
We first show that $\tau_{x(t)}=\min_{0\le r<m}T_r(t)$ for any $t \ge 0$. Indeed, let $h(t):=|\{i:x_i(t)>\tau_{x(t)}\}|\le m-1$. By the definition of $\tau$,
$$
\tau_{x(t)}=\frac{1}{m-h(t)}\sum_{i=h(t)+1}^n x_i^\downarrow(t) = T_{h(t)}(t).
$$
For $r<h(t)$, the first $h(t)-r$ terms of the tail sum are at least $\tau_{x(t)}$, hence $T_r(t)\ge\tau_{x(t)}$. For $r>h(t)$, removing $r-h(t)$ terms of size at most $\tau_{x(t)}$ likewise gives $T_r(t)\ge\tau_{x(t)}$. This proves the identity.\\

\noindent Now, $\sum_{i=r+1}^n x_i^\downarrow(t)$ is the sum of the $n-r$ smallest coordinates of $x(t)$. Increasing one coordinate by $\delta\ge0$ increases this sum by an amount between $0$ and $\delta$. Consequently,
$$
0\le T_r(t+\delta)-T_r(t)
\le\frac{\delta}{m-r}\le\delta.
$$
We get $0 \le \tau_{x(t+\delta)}-\tau_{x(t)} \le \delta$ because $\tau_{x(t+\delta)}-\tau_{x(t)} = T_{h(t+\delta)}(t+\delta) - T_{h(t)}(t) \le T_{h(t)}(t+\delta) - T_{h(t)}(t) \le \delta$, and analogously,  $\tau_{x(t+\delta)}-\tau_{x(t)} \ge T_{h(t+\delta)}(t+\delta) - T_{h(t+\delta)}(t) \ge 0$.
In particular, $\tau_{x(t)}$ is non-decreasing and continuous in $t$.\\

\noindent It remains to prove the derivative formula. On any open interval where $H:=\{i:x_i(t)>\tau_{x(t)}\}$ is constant, the water-filling identity gives with $h=|H|$,
$$
\tau_{x(t)}=\frac{1}{m-h} \sum_{i\notin H}x_i(t),
\qquad
g(x(t))=\sum_{i\in H}\big(x_i(t)\big)^2+
(m-h)\Big(\frac{1}{m-h}\sum_{i\notin H}x_i(t)\Big)^2.
$$
Differentiation therefore yields
$$
\frac{d }{d t}g(x(t))=
\begin{cases}
2t, & j\in H,\\
2\tau_{x(t)}, & j\notin H.
\end{cases}
$$
Since $0\le\tau_{x(t+\delta)}-\tau_{x(t)}\le\delta$ for all $\delta\ge0$, the difference $t-\tau_{x(t)}$ is non-decreasing, while $x_l(t)-\tau_{x(t)}$ is non-increasing for $l\neq j$ because $x_l(t) = x_l$ is constant. Thus, each index changes its membership in $H$ at most once, and $H$ changes at most $n$ times.\\

\noindent On every interval where $H$ is constant, the derivative computed above equals $2\max\{t,\tau_{x(t)}\}$. Since $g(x(t))$ and this expression are both continuous in $t$ (by the continuity of $\tau_{x(t)}$), the one-sided derivatives
agree at the finitely many boundaries. Hence, $g(x(t))$ is continuously differentiable on $(0,\infty)$, with derivative $\frac{d}{dt}g(x(t)) = 2\max\{t,\tau_{x(t)}\}$.
\end{proof}

\begin{lemma}
\label{lemma:supermodularity}
    $f$ is supermodular.
\end{lemma}
\begin{proof}
    Let $S \subseteq T \subset N$ and $j \in N\setminus T$. Because the function $U \mapsto \sum_{k \in U}p_k^2$ is modular, it suffices to show that
    $$g(p_{S \cup \{j\}}) - g(p_S) \le g(p_{T \cup \{j\}}) - g(p_T).$$
    Denoting the $j$-th unit vector as $e_j$, we can rewrite for $U \in \{S,T\}$,
    \begin{equation*}
        g(p_{U \cup \{j\}}) - g(p_U) = g(p_U + p_{\{j\}}) - g(p_U) = \int_0^{p_j} \frac{d}{dt} g(p_U + t e_j)dt,
    \end{equation*}
    because $\frac{d}{dt} g(p_U + t e_j) = 2\max\{t,\tau_{p_U + t e_j}\}$ is continuous in $t$ by Lemma~\ref{lemma:water-filling}.
    We also know that $p_S \le p_T$, and thus $\frac{d}{dt} g(p_S + t e_j) \le \frac{d}{dt} g(p_T + t e_j)$ for all $t \in [0, p_j]$.
\end{proof}

Because $f$ is supermodular and $f(\emptyset)=0$, we can rewrite the LP relaxation \eqref{LP} as a supermodular LP via the variable transformation $X_j = p_j C_j$. 
This idea has previously appeared in the literature \cite{Goemans1996,Queyranne1993}.
By the optimality of Edmonds' greedy algorithm for supermodular LPs, after sorting the jobs by non-increasing Smith ratios, i.e., $\frac{w_1}{p_1} \ge \frac{w_2}{p_2} \ge \ldots \ge \frac{w_n}{p_n}$, an optimal LP solution can be computed in strongly polynomial time via the greedy solution
\begin{equation}
\label{equation:LP-optimum}
    X_j^\textup{LP} = p_jC_j^\textup{LP} = f([j]) - f([j-1]) = \frac{1}{2} p_j^2 + \frac{1}{2}\left[ g(p_{[j]}) - g(p_{[j-1]})\right].
\end{equation}

Similarly, the right-hand side of constraints~\eqref{stochastic water-filling inequalities} can also be shown to be supermodular.
However, because $\E[g(P_S)]$ depends on the marginal distributions of all processing time variables, computing the values of the LP optimum may be difficult in this case, and requires access to more distributional information than running WSEPT does.

\begin{corollary}
    $F$ is supermodular.
\end{corollary}
\begin{proof}
As before, it suffices to show for any $S \subseteq T \subseteq N \setminus \{j\}$ that
    $$\E[g(P_{S \cup \{j\}})] - \E[g(P_S)] \le \E[g(P_{T \cup \{j\}})] - \E[g(P_T)]$$
because the function $U \mapsto \mu^2(U) - \sigma^2(U)$ is modular.
By Lemma~\ref{lemma:supermodularity}, it holds for all $p_1,\ldots,p_n\ge0$, and thus for all realizations of $P_1,\ldots,P_n$, that $g(p_{S \cup \{j\}}) - g(p_S) \le g(p_{T \cup \{j\}}) - g(p_T)$.
Taking expectations completes the proof.
\end{proof}

\section{Discussion}
The analyses of Kawaguchi and Kyan~\cite{Kawaguchi1986}, Schwiegelshohn~\cite{Schwiegelshohn2011}, and Jäger and Skutella~\cite{Jaeger2018} work with the structure of worst-case instances and, as Jäger and Skutella point out, do not use a lower bound on the optimum. Our proof follows the pattern of LP-based analyses instead: it compares WSPT with the optimal value of a linear relaxation and reduces this comparison to an inequality for a single machine. The relaxation strengthens the inequalities~\eqref{oldLP} by replacing $p(S)^2/m$ with the water-filling function~$g$. %In this picture, the water level of the tight instance of Kawaguchi and Kyan is exactly $\sqrt2$, the parameter used in the proof.
 
The stochastic results show that the relaxation yields more than a new proof. The variances of the processing times enter the analysis only through the term $p^2(S)$, because $\E[P_j^2]\le(1+\Delta)\E[P_j]^2$; everything else carries over realization by realization (see Lemma~\ref{lemma:WSEPT}). Together with a second load inequality (\ref{eq:general_bound_on_load^2_no2}), this gives a guarantee of order $\sqrt{\Delta}$ for WSEPT, which matches the known lower bound up to a constant factor.
 
Some questions remain open. The relaxation gap of the new LP relaxation \eqref{LP} lies between $9/8$ and $\frac{1}{2}(1+\sqrt{2})$. For exponentially distributed processing times, the performance guarantee of WSEPT lies between $1.243$~\cite{Jagtenberg2011} and $4/3$. It is open whether WSEPT has a guarantee independent of $\Delta$ for a fixed number of machines~\cite{Jaeger2018}. Finally, it would be interesting to know whether water-filling LP relaxations will lead to further improvements in the study of min-sum scheduling and related problems.

\subsubsection*{Acknowledgments}
This research was partially funded by the German Research Foundation (DFG) under project number 277991500.

\subsubsection*{Declaration of AI Usage}
The contents of this manuscript are entirely the work of its human authors.
No generative AI has been used to conceive the ideas or mathematical arguments presented, nor to produce any part of the written presentation or exposition.
The authors used Grammarly (with generative AI features disabled) as a spell- and grammar-checker.
Anthropic's Claude Opus 5.5 and OpenAI's ChatGPT 5.6 were used in the review process of this manuscript. More precisely, they helped identify a handful of typos and some outdated or incomplete attributions in the references.  Claude Opus 5.5 was also used to prepare Figure~\ref{fig:waterfilling}.

\bibliographystyle{splncs04}
\bibliography{bibliography}

\appendix

\section{Lower Bound on the WSPT Gap \texorpdfstring{of \eqref{oldLP}}{of (1)}}
In the introduction, we claim that the LP relaxation \eqref{oldLP} is too weak to attain the bound by Kawaguchi and Kyan for WSPT, even with the additional constraints $C_j \ge p_j$.
This fact is folklore, though we are not aware of it ever having been published. 
For completeness, we provide an example that demonstrates this fact.

Consider an instance with $m=2$ machines and $n=3$ jobs with processing times $(1,1,4)$ and weights $(1,1,4)$.
Because these jobs are already indexed in WSPT order, WSPT attains completion times $C^{\rm WSPT}=(1,1,5)$ and total weighted completion time $\sum w_j C_j^{\rm WSPT}=22$.
It is easy to verify that the completion-time vector $C=(1,1,4)$ satisfies the constraints of \eqref{oldLP} as well as the additional constraints, and that $\sum w_j C_j=18$. Therefore,
$$
  \frac{\sum w_j C_j^{\rm WSPT}}{\sum w_j C_j^{\rm LP}}
  \ge \frac{22}{18}
  >\frac{1}{2} \Big( 1+\sqrt2 \Big).
$$
Note that any order of jobs would be a valid WSPT order, and we have selected the worst one.
However, this does not weaken our argument because increasing the weights of the first two jobs by $2\epsilon$ and $\epsilon$, respectively, makes the choice of the WSPT order unique and attains a bound larger than $\frac{1}{2} ( 1+\sqrt2 )$.

\end{document}